\documentclass[aps,pre,twocolumn,superscriptaddress,showpacs,amsmath,amssymb,footinbib,longbibliography]{revtex4-1}
\usepackage[english]{babel}
\usepackage{amssymb,amsbsy,amsmath}
\usepackage{graphicx}
\usepackage{dcolumn}
\usepackage{bm}
\usepackage{subfigure}
\usepackage{color}
\usepackage{textcomp}
\usepackage[colorlinks,bookmarks=false,citecolor=blue,linkcolor=red,urlcolor=blue]{hyperref}

\newcommand{\op}[1]{\fontdimen12\textfont3=2pt\fontdimen12\scriptfont3=1.4pt\!\null\mathop{\protect\vphantom{#1}\smash{#1}}\limits_{\sim}\null\!}
\newcommand{\xref}[1]{\protect\ref{#1}}
\newcommand{\figref}[1]{Fig.~\protect\ref{#1}}
\newcommand{\fmref}[1]{(\protect\ref{#1})}
\def\bra#1{\langle \, {#1} \, | \,}
\def\ket#1{\, | \, {#1} \, \rangle}

\renewcommand{\eqref}[1]{Eq.~(\protect\ref{#1})}

\begin{document}
\title{Spin-phonon interaction induces tunnel splitting in
  single-molecule magnets}

\author{Kilian Irl\"ander}
\author{J\"urgen Schnack}
\email{jschnack@uni-bielefeld.de}
\affiliation{Fakult\"at f\"ur Physik, Universit\"at Bielefeld, Postfach 100131, D-33501 Bielefeld, Germany}

\date{\today}

\begin{abstract}
  Quantum tunneling of the magnetization is a major obstacle
  to the use of single-molecule magnets (SMMs) as basic
  constituents of next-generation storage devices.
  In this context, phonons are often only considered (perturbatively)
  as disturbances
  that promote the spin system to traverse the anisotropy barrier.
  Here, we demonstrate the ability of phonons to induce a tunnel splitting
  of the ground doublet which then reduces the required
  bistability due to Landau-Zener tunneling of the magnetization.
  Harmful are those phonons that modify the spin Hamiltonian so
  that its rotational symmetry about the field axis is destroyed.
  In our calculations we treat spins and phonons on the same footing by performing
  quantum calculations of a Hamiltonian where the single-anisotropy
  tensors are coupled to harmonic oscillators.
\end{abstract}


\maketitle

\section{Introduction}

\begin{figure}[ht!]
\centering
\includegraphics*[clip,width=0.69\columnwidth]{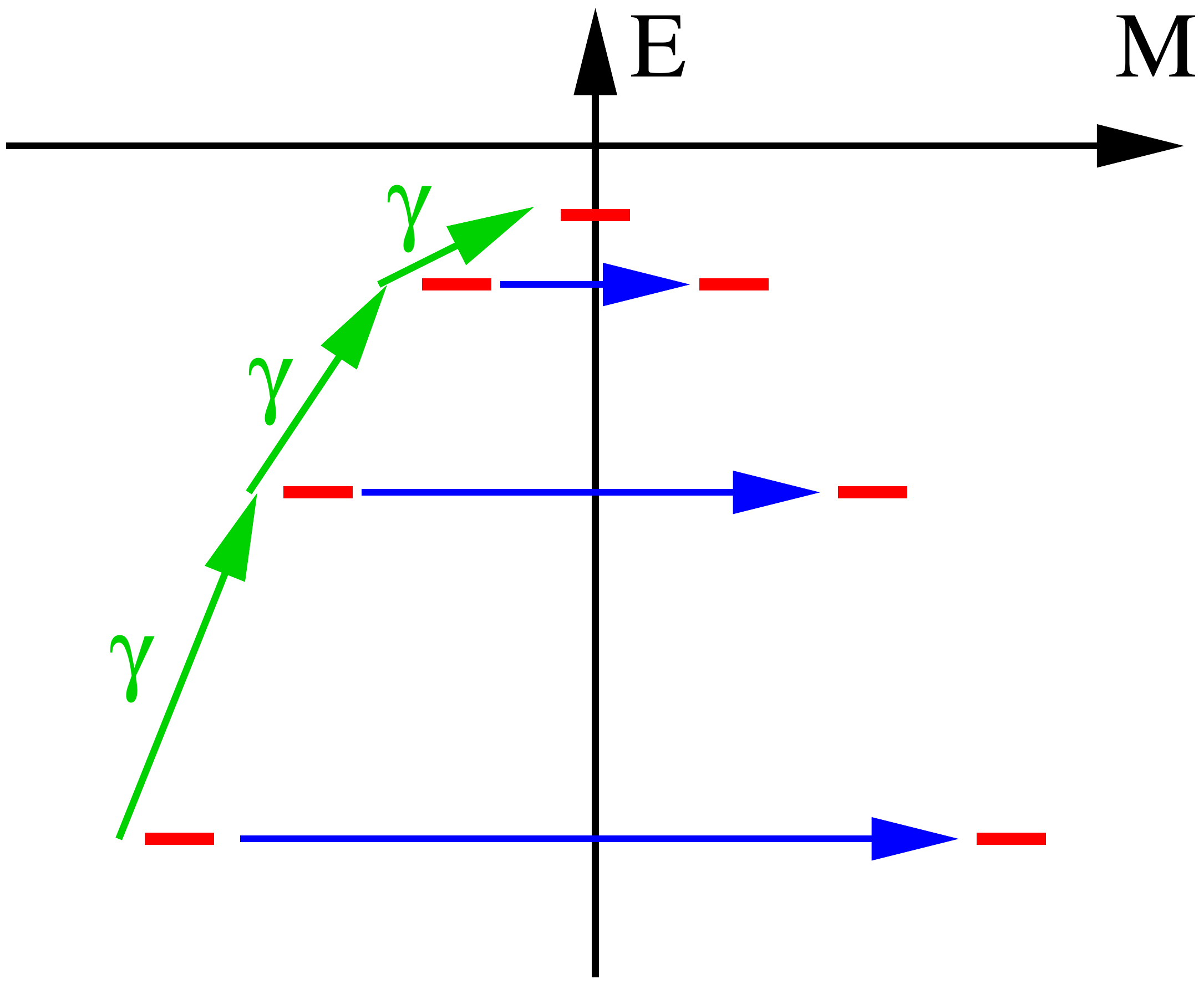}
\caption{Sketch of the low-lying energy levels of an SMM
  vs. magnetic quantum number.
  Red bars denote energy eigenvalues; they form an anisotropy
  barrier. The tiny (invisible) energy differences between
  almost degenerate states 
  for negative and positive magnetic quantum number,
  respectively, are 
  called tunnel splitting. Blue arrows show
  magnetization tunneling pathways for states with negative
  magnetic quantum number, and green arrows depict some of the
  possible 
  excitations due to phonons, compare e.g.\ \cite{LvS:CSR15}.} 
\label{spin-phonon-f-A}
\end{figure}

Single-molecule magnets (SMMs) constitute magnetic materials in
which the sufficiently separated molecules exhibit a magnetic
hysteresis of purely molecular origin. Typically,
such magnetic molecules are characterized by low-lying magnetic levels whose
energies form a barrier against magnetization reversal. This situation is
sketched in \figref{spin-phonon-f-A}. Initialized in one of
the two magnetic ground states, the system shows bistability and thus allows
the storage of information in the same way a bit on a hard drive
would do. Ever since the discovery of such behavior in Mn$_{12}$-acetate 
\cite{SGC:Nat93,FST:PRL96,TLB:Nature96,LTB:JAP97,CWM:PRL00,WSC:JAP00,GaS:ACIE03},
this promises future miniaturization of magnetic storage devices.

Among others, two major processes prevent an easy use of SMMs: quantum tunneling
of the magnetization between nearly degenerate levels at avoided level crossings,
depicted by blue arrows in \figref{spin-phonon-f-A},
as well as thermal excitations across the barrier transmitted by the phonons of
the molecule or the lattice of the crystal 
\cite{PhysRevB.61.1286,CGJ:PRL00,CGS:PRB05,TSN:IJQC05,TSY:P05,Blundell:CP07,PhysRevB.82.134403,Gla:CC11,LvS:CSR15,BFW:CEJ19,Sch:CP19}.
A very good and recent summary is provided in Ref.~\cite{LvS:CSR15}, where
particulary Figure 3 explains that the transitions imposed by direct, van Vleck,
Orbach, or second-order Raman processes are assumed to be of resonant nature
similar to scattering processes. Only recently this paradigm has been challenged by the
observation of sub-barrier tunneling which was subsequently modeled by means
of anharmonic phonons \cite{LTS:NC17}.
The question, what is or is not required to make a good SMM \cite{Nee:FD11},
got a new twist.

In the present article, we do not want to contribute to the discussion
of the role of the anisotropy barrier or the various transitions harmonic
or anharmonic phonons can induce
\cite{Wal:IC07,RCC:CC08,FHH:JACS10,BAJ:NC10,Gla:CC11,PDW:IC15,GOR:N17,EBG:CS18,ReC:PCCP19,NAB:JACS19,LuS:SA19,ABT:IC19}. 
Instead, we want to contribute a new question, namely whether
spin-phonon interactions can open a tunneling gap 
and thus enable temperature-independent groundstate tunneling of the magnetization.

\begin{figure}[ht!]
\centering
\includegraphics*[clip,width=0.69\columnwidth]{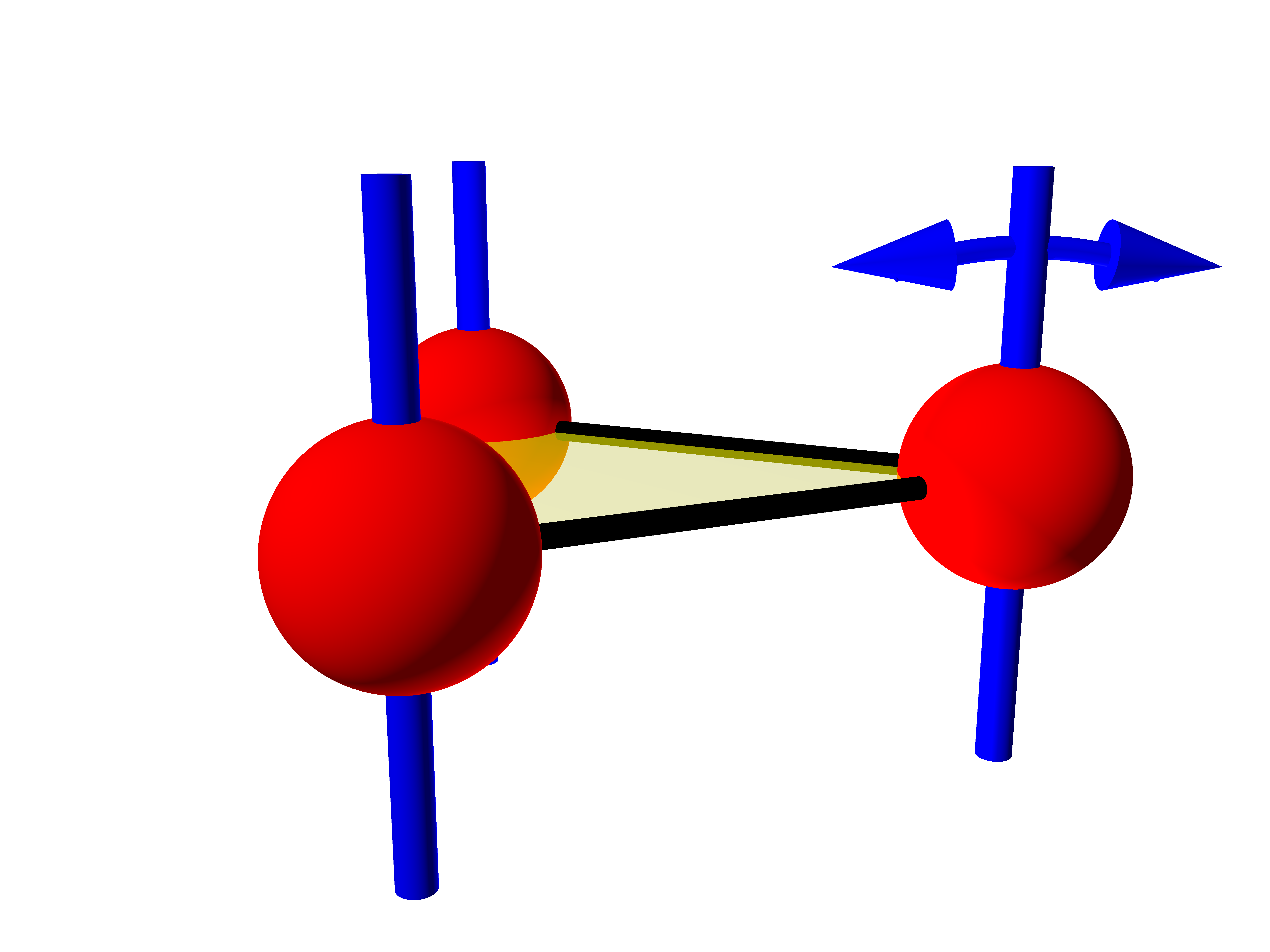}
\caption{Model system investigated in this article: red spheres
  denote spins that interact ferromagnetically with their
  neighbors, blue sticks represent single-ion easy-axis
  anisotropies. The latter are allowed to vibrate.}
\label{spin-phonon-f-B}
\end{figure}

In order to answer this question, we are going to investigate the model
system shown in \figref{spin-phonon-f-B}. Three spins at the corners of
a triangle interact with each other ferromagnetically. In addition, each
spin is subject to an easy-axis anisotropy. In our model, all three
anisotropy axes are coupled to harmonic oscillators and can thus fluctuate
away from their uniaxial rest direction. Such a simple, but yet realistic
model leads directly to the opening of a tunneling gap. Consequently, a
unitary time evolution of a state initialized at negative magnetic field
results in magnetization tunneling at zero field due to Landau-Zener transitions
\cite{WSC:JAP00}.

The paper is organized as follows. In Section \ref{sec-2}, we introduce the model
together with our numerical procedures. We present our numerical results in
Section~\ref{sec-3}. The article closes
with a discussion in Section~\ref{sec-4}.

\section{Method}
\label{sec-2}

The central idea of this paper is to investigate a combined
quantum system of spins and phonons by numerically exactly
diagonalizing an appropriate Hamiltonian.
For the purpose of demonstrating the ability of phonons to open
a tunneling gap, we choose a 
scenario where the  easy anisotropy axes are coupled to
vibrations of the molecule \cite{LTS:CS17}. As
shown schematically in \figref{spin-phonon-f-B}, our system
consists of three 
spins $s=1$ which couple ferromagnetically with each other and
are subject to a single-ion easy-axis anisotropy. Each of the
anisotropy axes is coupled to its own harmonic oscillator and
thus vibrates independently of the other axes. Thus, the total
Hamiltonian of our model consists of four parts, a Heisenberg
term, the single-ion anisotropy, the harmonic oscillators, and
the Zeeman term:
\begin{eqnarray}
\label{E-2-1}
\op{H}
&=&
-2 J\;
\left(\vec{\op{s}}_1\cdot \vec{\op{s}}_2 +\vec{\op{s}}_2\cdot \vec{\op{s}}_3 +\vec{\op{s}}_3\cdot \vec{\op{s}}_1 \right)
\\
&&
+
\vec{\op{s}}_1\cdot \mathbf{D}_1(\op{\theta}_1) \cdot \vec{\op{s}}_1
+
\vec{\op{s}}_2\cdot \mathbf{D}_2(\op{\theta}_2) \cdot \vec{\op{s}}_2
\nonumber
\\
&&
+
\vec{\op{s}}_3\cdot \mathbf{D}_3(\op{\theta}_3) \cdot \vec{\op{s}}_3
\nonumber
\\
&&
+
\omega_1\left(\op{a}^{\dagger}_1\op{a}_1^{\mathstrut}+\frac{1}{2}\right)
+
\omega_2\left(\op{a}_2^{\dagger}\op{a}_2^{\mathstrut}+\frac{1}{2}\right)
\nonumber
\\
&&
+
\omega_3\left(\op{a}^{\dagger}_3\op{a}_3^{\mathstrut}+\frac{1}{2}\right)
\nonumber
\\
&&
+
g\mu_B\cdot \vec{B}\cdot \left( \vec{\op{s}}_1+\vec{\op{s}}_2+\vec{\op{s}}_3\right)
\nonumber
\ .
\end{eqnarray}
Here, $\vec{\op{s}}_i$ are the spin-vector operators of the three
spins; $J>0$ denotes the ferromagnetic
coupling. The $\mathbf{D}_i(\op{\theta}_i)$ model the easy
anisotropy axes
\begin{eqnarray}
\label{E-2-2}
\mathbf{D}_i(\op{\theta}_i)
&=&
D\;
\vec{e}_i(\op{\theta}_i,\phi_i) \otimes
\vec{e}_i(\op{\theta}_i,\phi_i)
\\
\vec{e}_i(\op{\theta}_i,\phi_i)
&=&
\begin{pmatrix}
  \cos(\phi_i)\sin(\op{\theta}_i)\\
  \sin(\phi_i)\sin(\op{\theta}_i)\\
  \cos(\op{\theta}_i)\\
\end{pmatrix}
\\
\phi_i &=& (i-1) 2\pi/3,\qquad
i\in\{1,2,3\}
\ .
\end{eqnarray}
They depend on two angular coordinates of which
$\op{\theta}_i$ couples to the harmonic oscillator degrees of
freedom
\begin{eqnarray}
\label{E-2-3}
\op{\theta}_i
&=&
\theta_{i,0}+\alpha\sqrt{\omega}\op{x}_i,\qquad
i\in\{1,2,3\}
\\
&&
\op{x}_i
\propto
\left(\op{a}^{\dagger}_i+\op{a}_i^{\mathstrut}\right)
\ .
\end{eqnarray}
We assume the same coupling strength $\alpha$ for all three
anisotropy axes with their respective oscillators and
the same frequency $\omega$ for all of these oscillators.
It is obvious that this model can be easily generalized to less 
idealized cases. 

Without loss of generality, we choose a ferromagnetic exchange of
$J=10$~K and an easy anisotropy of $D=-5$~K for the following
calculations. The external field will always point in global
$z$-direction, i.e.\ $\vec{B}=B\vec{e}_z$.

For numerical diagonalization, we have to represent Hamiltonian
\fmref{E-2-1} with respect to a basis. We use the product basis
$\left\{ \ket{m_1, m_2, m_3; n_1, n_2, n_3}  \right\}$, where
$m_i=-1,0,1$ are the magnetic quantum numbers of the individual
spins and $n_i=0,1,\dots$ are the quantum numbers of the three
oscillators. Matrix elements containing sine or cosine functions
of $\op{\theta}_i$ can be evaluated exactly with the help of an
intermediate basis transform. The dimension of the underlying
Hilbert space is 
infinite due to the harmonic oscillators. We therefore cut the
oscillator quantum numbers at some $n_{\text{max}}$ for our
numerical treatment. We investigated carefully that this indeed
does not influence the qualitative conclusions of the present
paper. Under these conditions, it turns out that
$n_{\text{max}}=1$ is sufficient to demonstrate our main
finding, the opening of a tunneling gap.

The numerical solutions of the stationary as well as of the
time-dependent Schr\"odinger equation presented in the next
section have been obtained with
\emph{Mathematica}\texttrademark\ 
in a recent thesis \cite{Irl:Master20}.

\section{Numerical results}
\label{sec-3}

The interesting case is given by a situation where the anisotropy
axes are at rest and point uniaxially along the field direction, which
is chosen in positive global $z$-direction. This means $\theta_{i,0}=0,
\forall i$. Without phonons, such a system would possess a
perfect crossing of the two degenerate ground state levels at
$B=0$, and thus would not show any magnetization tunneling. 

\begin{figure}[ht!]
\centering
\includegraphics[width=0.69\columnwidth]{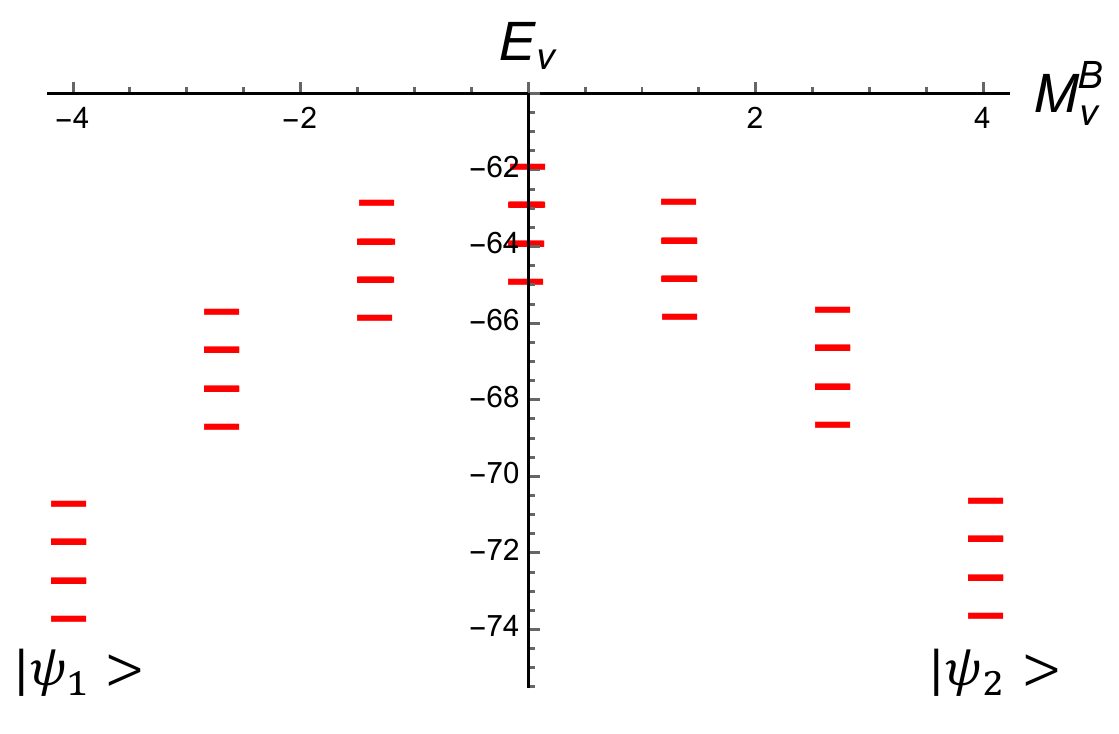}

\includegraphics[width=0.69\columnwidth]{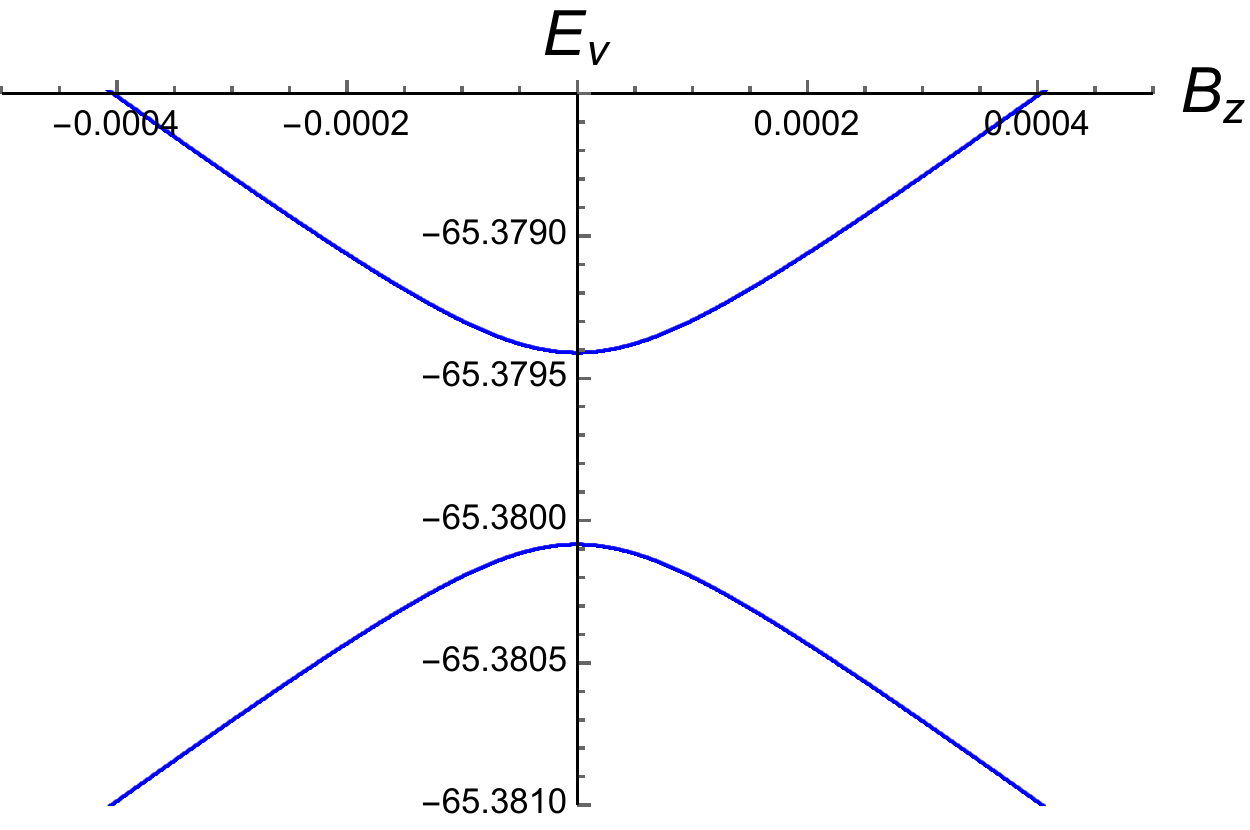}
\caption{Top: Energy levels vs. magnetic moment
  $\mathcal{M}_\nu^B=-g\mu_B\bra{\nu}\op{S}^z\ket{\nu}$ for
  $\alpha=0.01$, $n_{\text{max}}=1$, $\omega=1$. The two nearly
  degenerate ground states are denotes as $\ket{\Psi_1}$ and
  $\ket{\Psi_2}$. 
  Bottom: Avoided level crossing of the lowest doublet for
  $\alpha=0.5$. The
  energy difference at $B=0$ is the tunnel splitting $E_G$.
}
  \label{spin-phonon-f-C}
\end{figure}

If the coupling between spins and phonons is switched on, one
observes two results. Each level of the spin Hamiltonian splits
into a bunch of combined spin-phonon levels, and level crossings,
e.g.\ of the ground state, turn into avoided level crossings. 
The width of a bunch, compare top of \figref{spin-phonon-f-C},
depends both on $\alpha$ and $\omega$.
The number of levels in a bunch is given by the dimension of the
phonon subspace, in our calculation 8 (but the levels are 
degenerate). The bottom of \figref{spin-phonon-f-C} shows the
behavior of the lowest levels at the avoided level crossing.
The energy difference at $B=0$ is the tunnel splitting $E_G$

\begin{figure}[ht!]
\centering
\includegraphics[width=0.69\columnwidth]{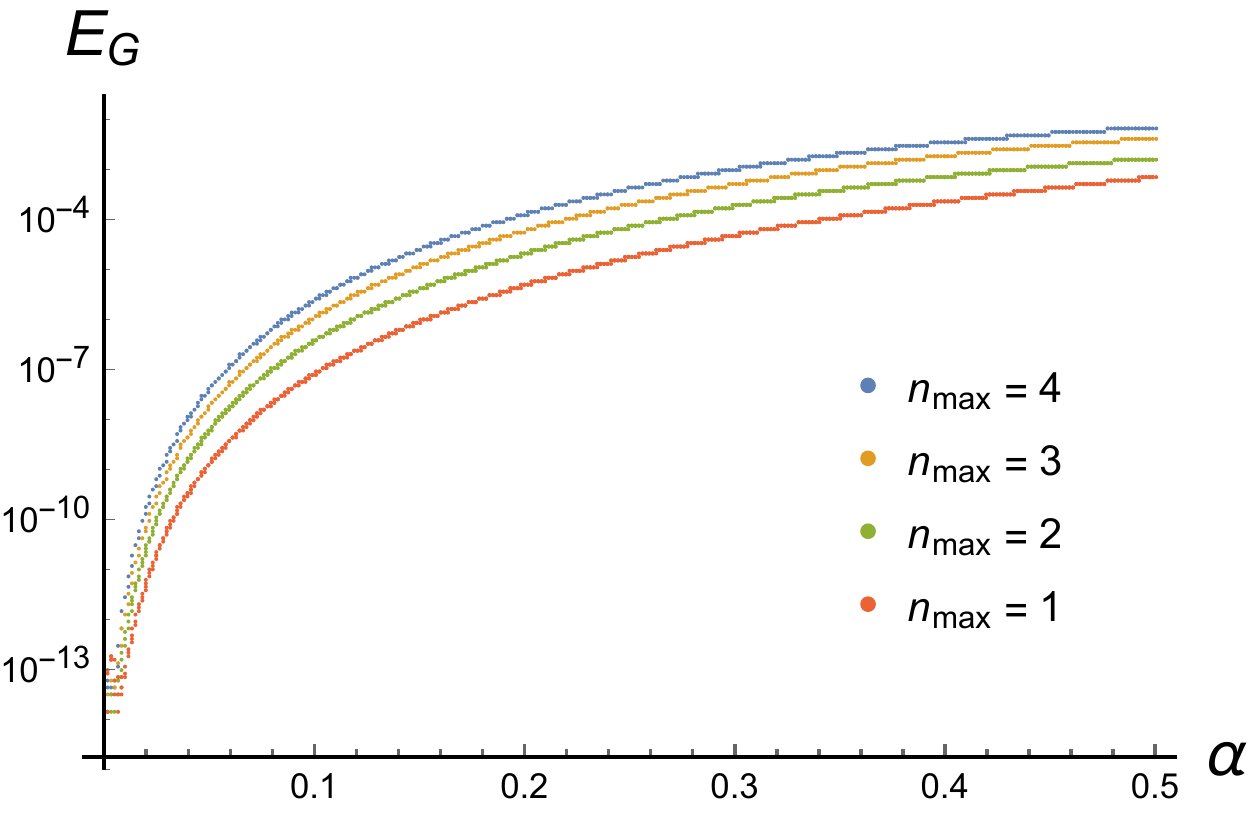}

\includegraphics[width=0.69\columnwidth]{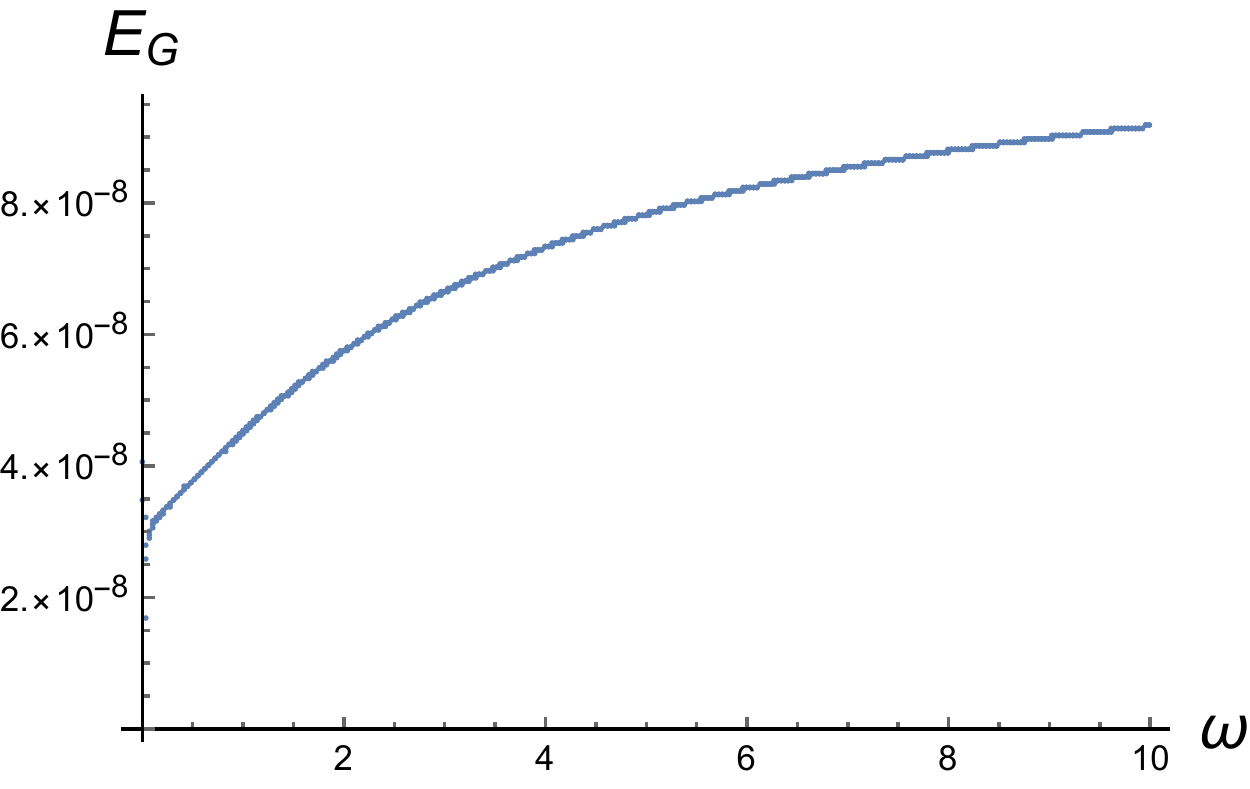}
\caption{Size of the tunnel splitting $E_G$. Top: $E_G$ vs.
  $\alpha$ for $\omega=5$ and $n_{\text{max}}\in\{1,2,3,4\}$.
  Bottom:  $E_G$  vs. $\omega$ for $\alpha=0.1$ and
  $n_{\text{max}}=1$.}
  \label{spin-phonon-f-D}
\end{figure}

The size of the tunneling gap $E_G$ depends strongly on $\alpha$ and
rather weakly on $\omega$, as can be seen in
\figref{spin-phonon-f-D}. For values $E_G<10^{-13}$, the
numerical determination of the tunneling gap is no longer
accurate. For values of $\alpha$ in the range from 0.02 to 0.5,
the values of the 
tunneling gap $E_G$ span nine orders of magnitude. This means
that vibrations of the anisotropy tensors can have a massive
influence on the character of the low-lying energy spectrum.
This tendency increases with the number of available phonons,
i.e.\ the number of oscillator excitations. A larger number of
available oscillator excitations corresponds to a larger quantum
mechanical (as well as thermal) variance of the deflection of
all anisotropy tensors.

Finally, we calculate the time evolution for an initial state
for various sweep rates $\dot{B}$ of the external magnetic
field. For our calculations, we choose $\alpha=0.5$, $\omega=5$
and $n_{\text{max}}=1$. This results in a relatively strong
coupling between oscillators and anisotropy axes, but we choose
this for educational reasons since the effect is
stronger and therefore easier to observe. For
smaller couplings $\alpha$, it will of course also be present,
just less obvious.

\begin{figure}[ht!]
\centering
\includegraphics[width=0.65\columnwidth]{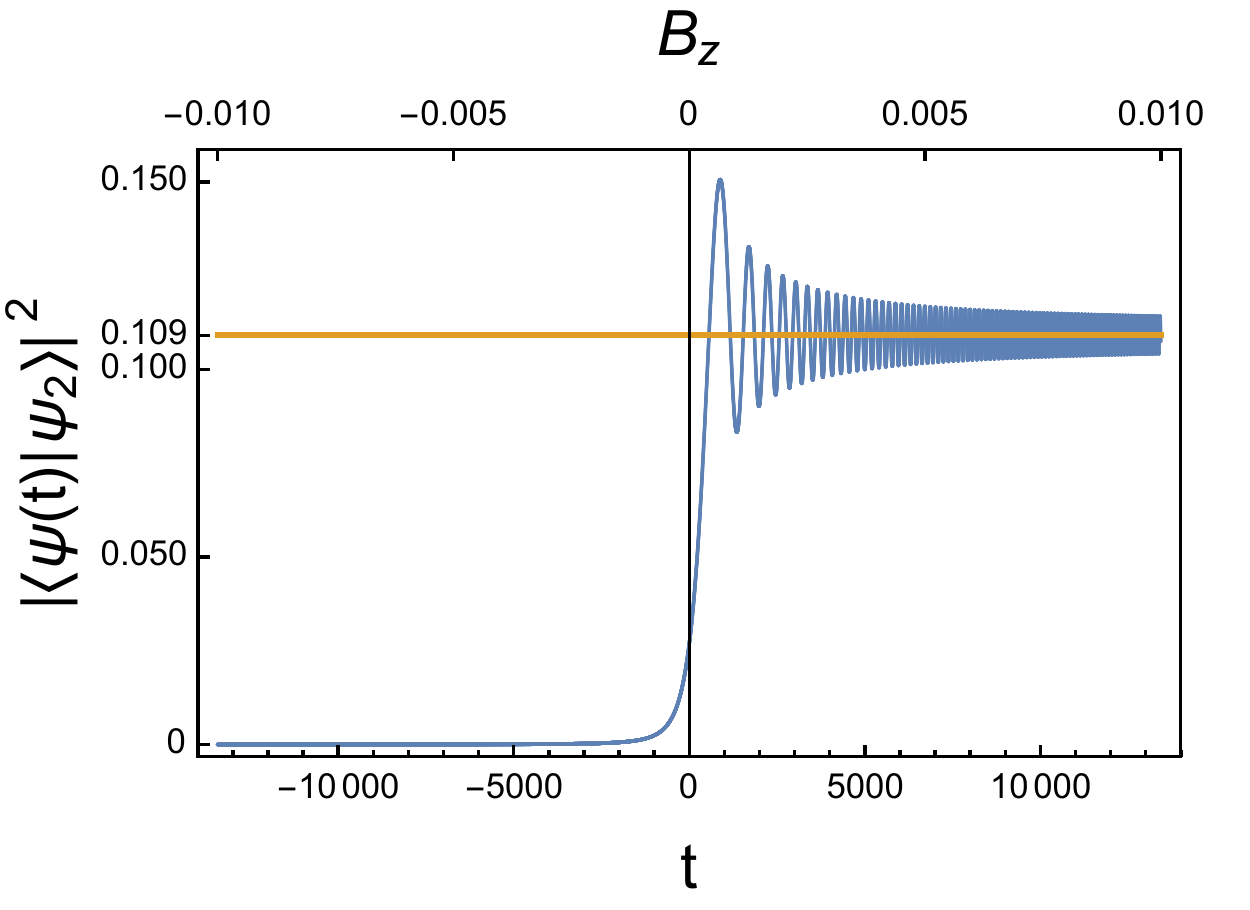}

\includegraphics[width=0.65\columnwidth]{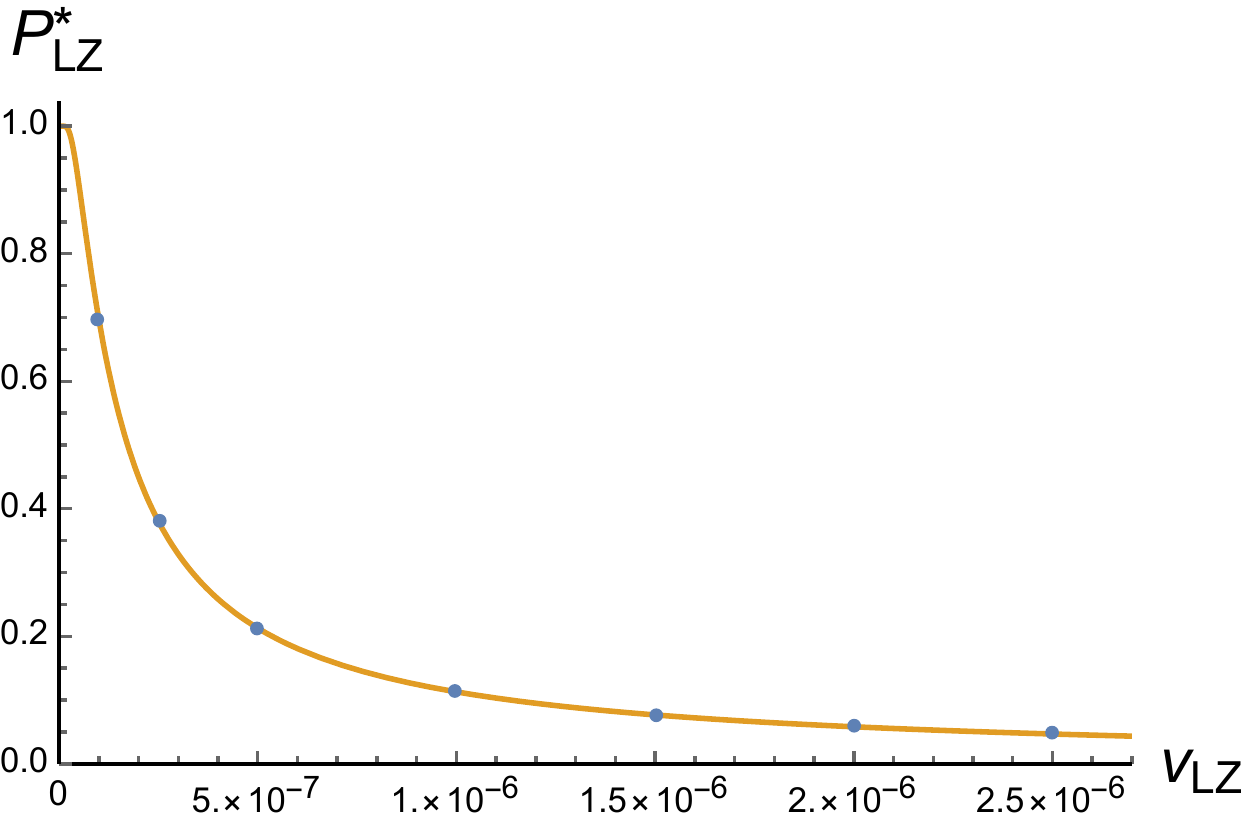}
\caption{Landau-Zener transitions in the case
  of strong coupling $\alpha=0.5$ and $\omega=5$ for
  $n_{\text{max}}=1$. The initial state is $\ket{\Psi_1}$.
  Top: Example of the unitary time evolution with
  $\dot{B}=1\cdot 10^{-6}$.
  Bottom: Transition probability at the
  avoided crossing of the two ground state levels as a function
of $\dot{B}$. Blue dots are numerical integrations, the solid
curve corresponds to the Landau-Zener theory.} 
  \label{spin-phonon-f-E}
\end{figure}

Figure~\xref{spin-phonon-f-E} shows the time evolution of the
overlap between the time-evolved state $\ket{\Psi(t)}$ and the
asymptotic state $\ket{\Psi_2}$ the system tunnels into for a certain
sweep rate $\dot{B}$ of the external field (top).
The time evolution starts with
$\ket{\Psi(t_{\text{start}}})=\ket{\Psi_1}$, 
see \figref{spin-phonon-f-C},
at some initial time and negative external field.
While passing $B=0$, the system undergoes a
Landau-Zener transition and oscillates about a limiting overlap
with the state $\ket{\Psi_2}$, compare
\figref{spin-phonon-f-E}. Such time evolutions have been
performed for several sweep rates. The resulting transition
rates, i.e.\ the probabilities to find the system in the other
limiting state, are given by blue dots in
\figref{spin-phonon-f-E}~(bottom). These numerical values
coincide perfectly with the results obtained from Landau-Zener
theory which also tells us that the higher-lying excited states of the
system do not really mix in at the transition. With that
confidence, one can now read off the transition probabilities for
smaller sweep rates that are not accessible in numerical
time evolutions due to the prohibitively large number of time
steps one would have to perform.

\begin{figure}[ht!]
\centering
\includegraphics[width=0.65\columnwidth]{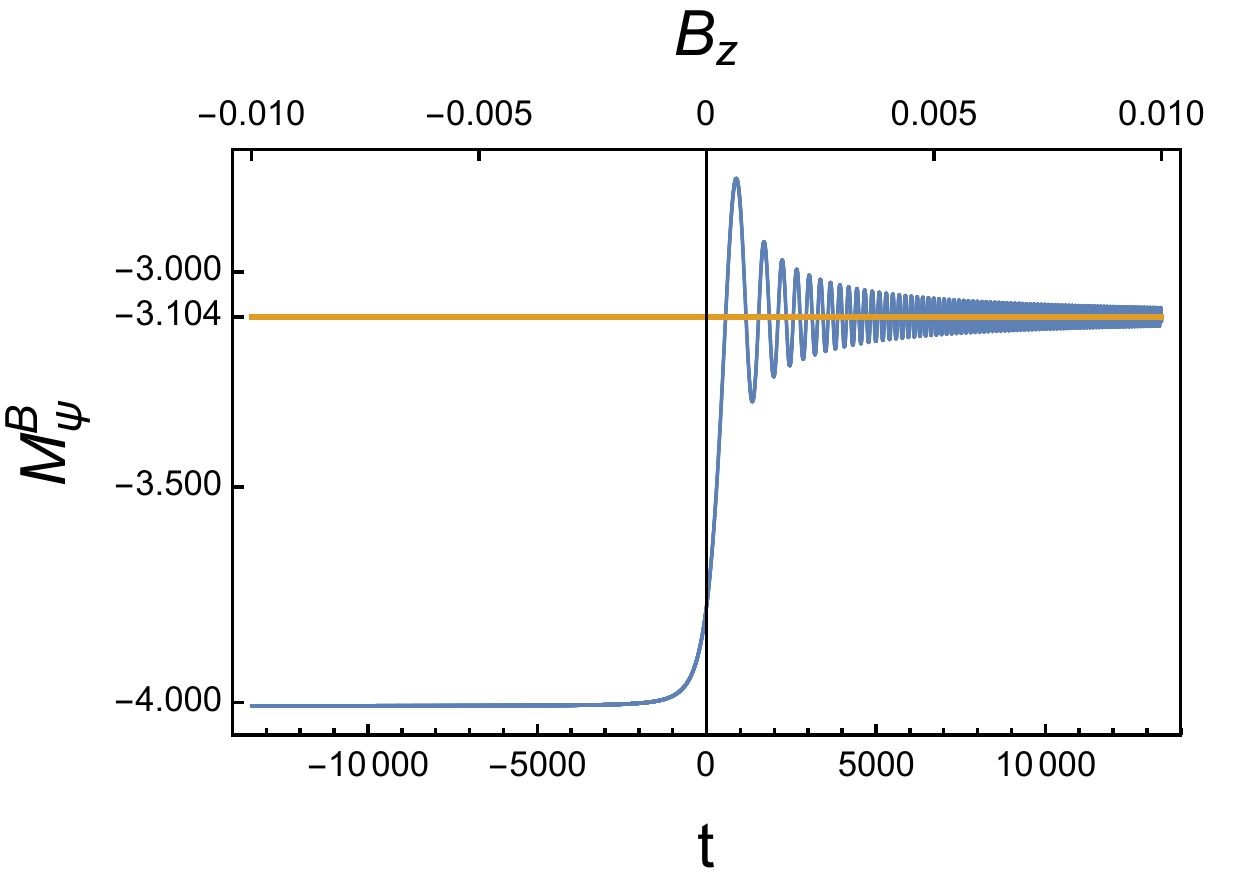}

\includegraphics[width=0.65\columnwidth]{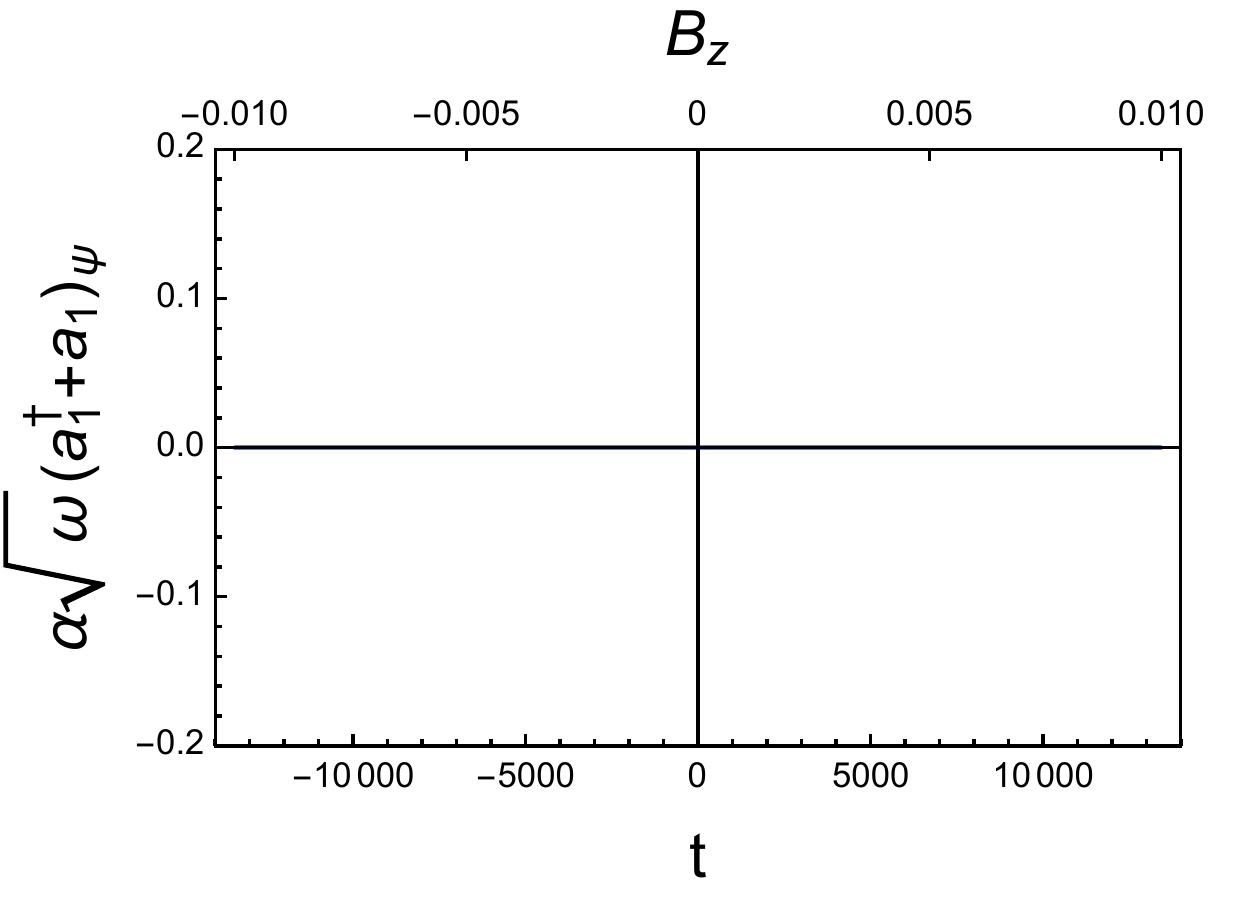}
\caption{Landau-Zener transitions in the case
  of strong coupling $\alpha=0.5$ and $\omega=5$ for
  $n_{\text{max}}=1$. The initial state is $\ket{\Psi_1}$.
  Top: Magnetization tunneling during the unitary time evolution with
  $\dot{B}=1\cdot 10^{-6}$.
  Bottom: Expectation values of
  $\alpha\sqrt{\omega}(\op{a}^{\dagger}_1+\op{a}_1)\propto\op{\theta}_1$.
  $\op{\theta}_2$ and $\op{\theta}_3$ behave in the same
  way. All numerical values are virtually zero.}   
  \label{spin-phonon-f-F}
\end{figure}

Finally, we want to discuss the behavior of the phonon
subsystem, i.e.\ the dynamics of the anisotropy axes. For the
time evolution depicted in \figref{spin-phonon-f-E}, where one
of the two nearly degenerate ground states is taken as initial
state and evolved via a time-dependent magnetic field, we can
not only evaluate the amount of magnetization that tunnels. We
can also calculate the phononic excitations, i.e.\ the
vibrations of the anisotropy axes while the system is swept
across $B=0$. Both $\ket{\Psi_1}$ and $\ket{\Psi_2}$ are practically in
their phononic ground states, i.e.\ the phononic contribution to
these eigenstates of the total Hamiltonian is dominantly
$\ket{n_1=0, n_2=0, n_3=0}$. It turns out that this does not really change
in the course of the time evolution, i.e.\ phonons are not
excited and the anisotropy axes do not vibrate noticibly,
nevertheless some part of the magnetization tunnels,
compare Figure~\xref{spin-phonon-f-F}.

\begin{figure}[ht!]
\centering
\includegraphics[width=0.65\columnwidth]{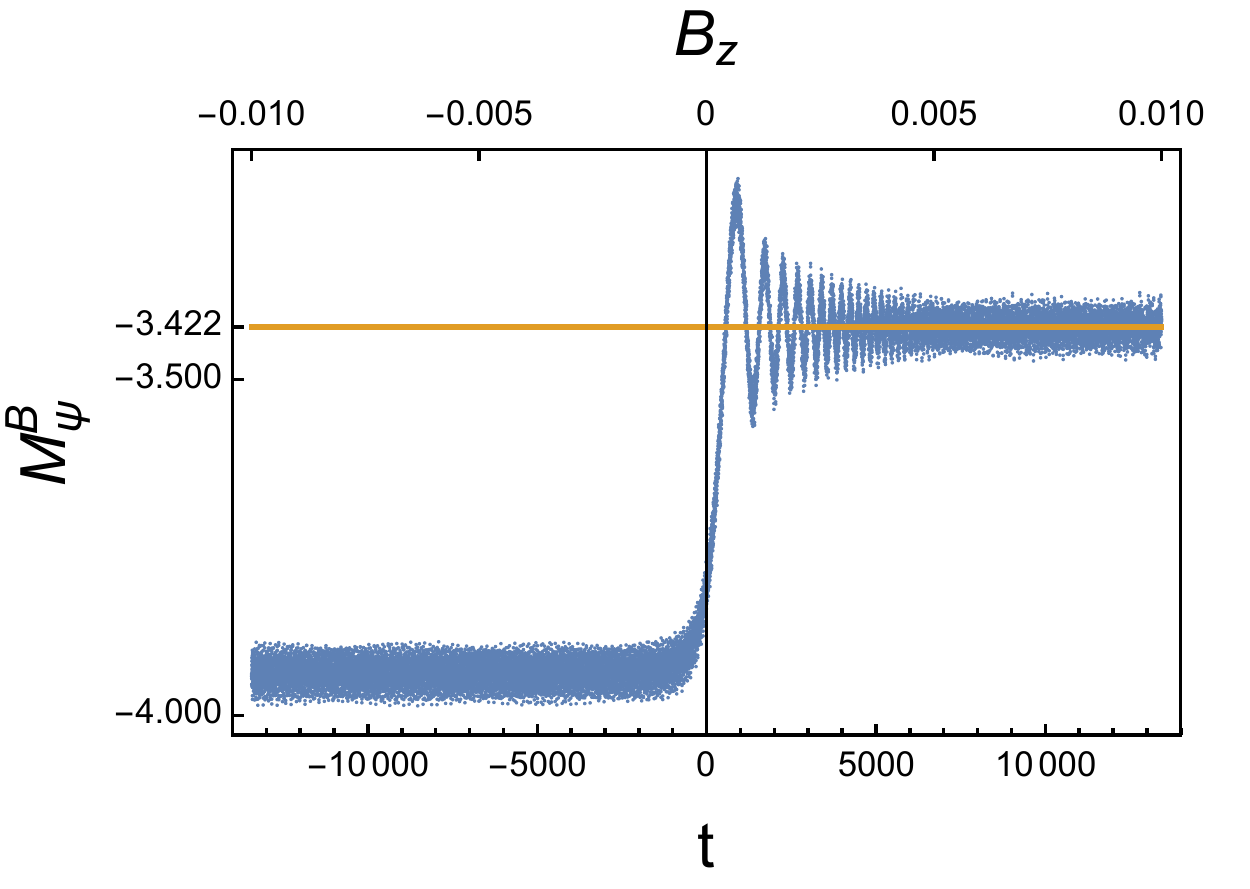}

\includegraphics[width=0.65\columnwidth]{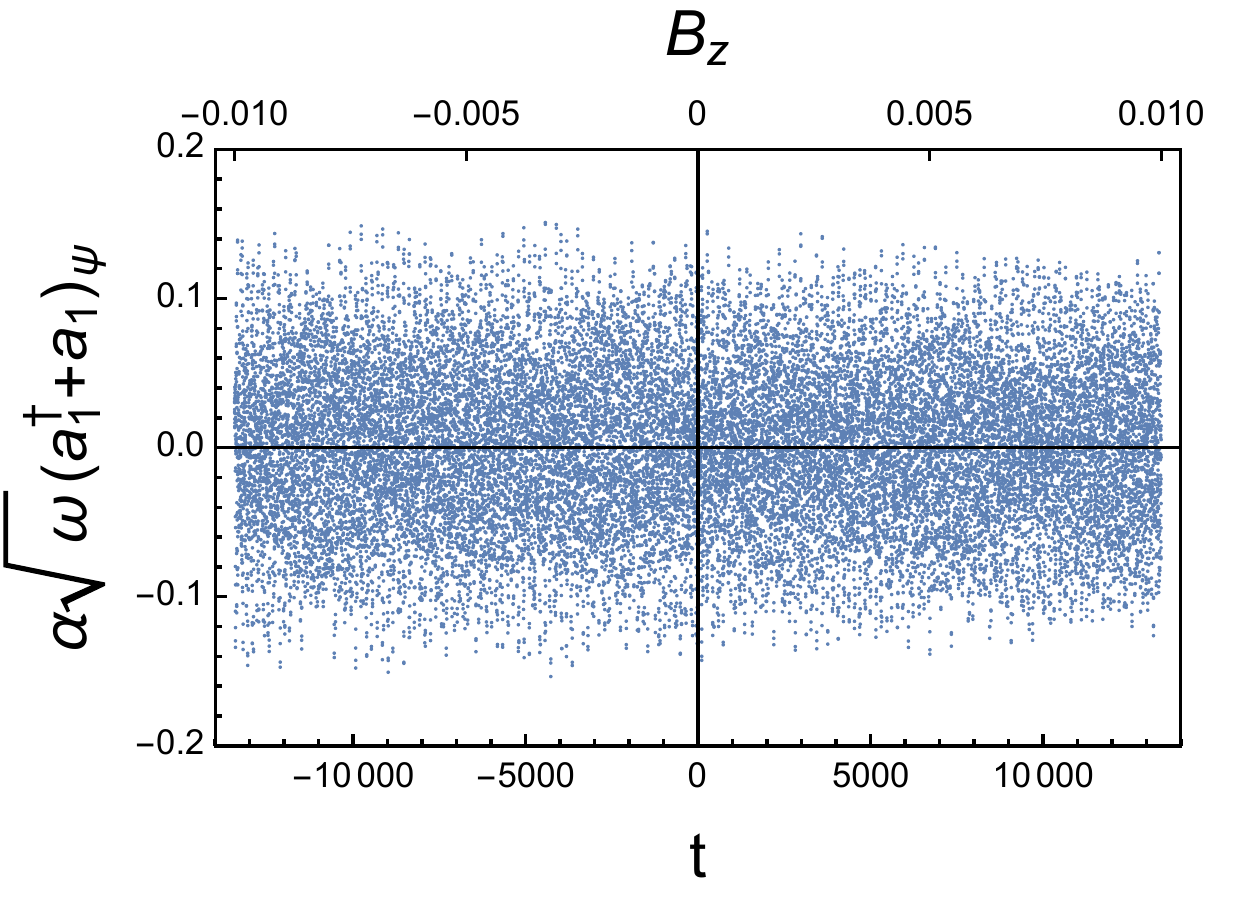}
\caption{Landau-Zener transitions in the case
  of strong coupling $\alpha=0.5$ and $\omega=0.1$ for
  $n_{\text{max}}=1$. The initial state is a supersosition
  of all state in the bunch of $\ket{\Psi_1}$, see \figref{spin-phonon-f-C}.
  Top: Magnetization tunneling during the unitary time evolution with
  $\dot{B}=1\cdot 10^{-6}$. 
  Bottom: Expectation values of
  $\alpha\sqrt{\omega}(\op{a}^{\dagger}_1+\op{a}_1)\propto\op{\theta}_1$. $\op{\theta}_2$
  and $\op{\theta}_3$ behave in 
  the same way. The dynamics contains very high frequencies due
  to which the plots look fuzzy. Each plot shows 20,000 points
  which corresponds to $\Delta t\approx 1$ between
  points. \emph{Mathematica}\texttrademark\ uses a variable step
  size that is much smaller.
} 
  \label{spin-phonon-f-G}
\end{figure}

We also considered a superposition of initial states that belong
to a bunch of levels which may be simultaneously occupied at
non-zero temperature. For this purpose we chose a smaller
$\omega=0.1$. The situation is displayed in
\figref{spin-phonon-f-G}. Although the anisotropy axes vibrate
fiercely about the $z$-direction with a high internal frequency
and a much larger amplitude, the effect on the magnetization
tunneling is only slightly 
bigger compared to a situation with only $\ket{\Psi_1}$ as
initial state but adapted smaller $\omega=0.1$ (not shown).
We suppose that the internal frequency of the phonon subsystem
is just too big in the present 
scenario, so that the spin system is too inert to follow.

The last investigations lead us to the final and important
conclusion that the mere coupling $\alpha$ of the spins to the
zero-point motion of the oscillator/phonon subsystem is 
sufficient to open the tunneling gap and induce Landau-Zener
transitions. Although the expectation value of the anisotropy
axes does not deviate from the uniaxial $z$-direction, the fact
that this value is not sharp but subject to a quantum mechanical
variance leads to the reduction of bistability.

\section{Discussion and conclusions}
\label{sec-4}

In this article, we were able to demonstrate that phonons can
open up a tunneling gap between otherwise degenerate ground
states of a single-molecule magnet. In our model calculation, the
phonons destroyed the uniaxial character of the single-ion
easy-axis anisotropies. We expect that phonons breaking the
rotational symmetry about the common axis of the field and the
anisotropy tensors in any other way would result in a similar
effect.

Our investigations are in line with earlier studies of
spin-phonon interactions, for instance in
Ref.~\cite{ElM:EPJB05}, where the phonons that modify the
exchange integrals are treated classically as well as in
Ref.~\cite{LRP:PRB09} where a distortion-dependent
Dzyaloshinskii-Moriya interaction is considered. Despite having
a different focus, both papers agree with our observation that
if the spin-phonon coupling is too strong, static distortions of
the system occur since the combined system of spins and
lattice becomes spin-Peierls unstable. 

Summarizing, we would like to state that investigations such as
the present one help determining prerequisites for the
successful design of single-molecule magnets by identifying
harmful vibrations that should be suppressed in such molecules.

\section*{Acknowledgment}

This work was supported by the Deutsche Forschungsgemeinschaft DFG
(314331397 (SCHN 615/23-1); 355031190 
(FOR~2692); 397300368 (SCHN~615/25-1)). We thank Patrick
Vorndamme for carefully reading the manuscript.



%

\end{document}